\documentclass{elsart}
 \usepackage{amssymb}
 \usepackage{epsfig}
    \newcommand{\ba}{\begin{eqnarray}}
    \newcommand{\ea}{\end{eqnarray}}
    \newcommand{\be}{\begin{equation}}
    \newcommand{\ee}{\end{equation}}

    \newcommand {\bk} {{\mathbf k}}
    \newcommand {\bn} {{\mathbf n}}
    
    \newcommand {\bx} {{\mathbf x}}
    \newcommand {\by} {{\mathbf y}}
    
    \newcommand {\calM} {{\mathcal M}}
    
    \newcommand{\AmS}{{\protect\the\textfont2
  A\kern-.1667em\lower.5ex\hbox{M}\kern-.125emS}}
  \newcommand {\calF}{{\mathcal F}}

\begin{document}
\runauthor{Junhua and Chuan}
\begin{frontmatter}

\title{A Numerical Study of Improved Quark Actions on
Anisotropic Lattices\thanksref{fund}}
\author[Beida]{Shiquan Su},
\author[Beida]{Liuming Liu},
\author[Beida]{Xin Li}
\author[Beida]{and Chuan Liu}
\address[Beida]{School of Physics\\
          Peking University\\
          Beijing, 100871, P.~R.~China}
\thanks[fund]{This work is supported by the Key Project of National Natural
Science Foundation of China (NSFC) under grant No. 10235040 and
supported by the Trans-century fund from Chinese
 Ministry of Education.}

 \begin{abstract}
 Tadpole improved Wilson quark actions with clover terms
 on anisotropic lattices are studied numerically.
 Using asymmetric lattice volumes, the
 pseudo-scalar meson dispersion relations are measured
 for $8$ lowest lattice momentum modes with quark mass values
 ranging from the strange to the charm quark with various values of
 the gauge coupling $\beta$ and $3$ different values of the bare speed of
 light parameter $\nu$. These results can be utilized to extrapolate or
 interpolate to obtain the optimal value for the bare speed of
 light parameter $\nu_{opt}(m)$ at a given gauge coupling
 for all bare quark mass values $m$. In particular, the optimal values
 of $\nu$ at the physical strange and charm quark mass
 are given for various gauge couplings.
 The lattice action with these optimized parameters can then
 be used to study physical properties of hadrons involving either
 light or heavy quarks.
 \end{abstract}
 \end{frontmatter}


\section{Introduction}

 It has become clear that anisotropic lattices and improved
 lattice actions are the ideal candidates for lattice
 QCD calculations involving heavy objects like the glueballs,
 one meson states with non-zero three momenta and
 multi-meson states with or without three momenta.
 It is also a good workplace for the study of hadrons
 with heavy quarks. In this work we present our numerical
 study on the quark action parameters suitable for
 heavy flavor physics. The gauge action employed in
 this paper is the tadpole improved
 gluonic action on asymmetric lattices:
 \ba
 S=&-&\beta\sum_{i>j} \left[
  {5\over 9}{TrP_{ij} \over \xi u^4_s}
 -{1\over 36}{TrR_{ij} \over \xi u^6_s}
 -{1\over 36}{TrR_{ji} \over \xi u^6_s} \right] \nonumber \\
 &-&\beta\sum_{i} \left[ {4\over 9}{\xi TrP_{0i} \over  u^2_s}
 -{1\over 36}{\xi TrR_{i0} \over u^4_s} \right] \;\;£¬
 \ea
 where $P_{ij}$ is the usual plaqette variable and
 $R_{ij}$ is the $2\times 1$ Wilson loop on the lattice.
 The parameter $u_s$, which we take to be the forth root
 of the average spatial plaquette value, incorporates the
 so-called tadpole improvement and $\xi$ designates the (bare)
 aspect ratio of the anisotropic lattice.
 With the tadpole improvement in place, the bare anisotropy
 parameter $\xi$ suffers only small renormalization which
 we neglect in this study.
 Using this action, glueball and light hadron spectrum has been studied
 within the quenched approximation \cite{colin97,colin99,%
 chuan01:gluea,chuan01:glueb,chuan01:canton1,chuan01:canton2,chuan01:india}.

 It has been suggested that relativistic heavy quarks
 can also be treated with the help of anisotropic lattices
 (the Fermi lab approach),
 possibly with improvements \cite{kronfeld97:aniso,%
 klassen98:wilson_quark,klassen99:aniso_wilson,kronfeld01:aniso_Oa,%
 harada02:aniso}.
 Using various versions of the quark actions, charmed meson
 spectrum, charmonium spectrum, charmed baryons have
 been studied on the
 lattice~\cite{mackenzie98:BsDs,chen01:aniso,lewis01:aniso,%
 CPPACS02:aniso,juettner03:Ds,nemeto03:aniso_baryon}.
 Another type of application of the anisotropic lattices is
 the calculation of hadron-hadron scattering lengths within the quenched
 approximation~\cite{chuan02:pipiI2,chuan04:KN,chuan04:Kpi,chuan04:pipi}.
 However, in order to take full advantage of the
 improved quark action on anisotropic lattices, some parameters
 in the action have to be determined, either perturbatively or
 non-perturbatively, in order to gain as much improvement as possible.
 Some numerical studies of these parameters have already appeared in
 the literature~\cite{chuan01:tune,harada02:aniso,CPPACS03:aniso_unquench,%
 okamoto03:aniso,peardon04:aniso}.
 The anisotropic quark actions used in these studies fall
 into two categories. These two cases differ mainly in the
 choice of spatial Wilson parameter $r_s$. According to the tree-level
 study~\cite{kronfeld97:aniso,kronfeld01:aniso_Oa},
 the choice of $r_s=1/\xi$ has a virtue
 that the optimal parameters in the action
 as a function of the quark mass contains no corrections of
 the form $m_0a_s$. The quark mass dependent corrections
 comes in only in terms of $m_0a_t$ which is assumed to
 be small. As a result, the optimal values of the parameters
 can be approximated by their values in the zero quark mass limit.
 That is to say, tuning of the parameters in the action becomes
 almost quark mass independent.
 The disadvantage of this choice
 is that the doubler are not very well separated from the
 ordinary fermions, particularly for large $\xi$. In the other
 choice, one sets $r_s=1$. This presumably elevates the doublers
 well above the ordinary fermion modes, however, the optimal
 values of the parameters
 in this choice might receive $O(m_0a_s)$ corrections, as
 suggested by tree-level and one-loop perturbative
 studies~\cite{kronfeld97:aniso,kronfeld01:aniso_Oa}.
 Therefore, if one takes the choice of $r_s=1$, optimal values
 of the action parameters in principle must be tuned in a
 quark {\em mass-dependent} way.

 In this paper, we will discuss the tuning of
 the bare speed of light parameter $\nu$
 in a quenched calculation using tadpole-improved Wilson
 fermions on anisotropic lattices.
 The parameter $\nu$ has to be tuned such that the
 pseudo-scalar meson energy-momentum dispersion relation
 reproduces its continuum form in the low-momentum limit.
 The dispersion relations of pseudo-scalar mesons
 are measured in our simulation for quark mass values ranging
 from the strange to the charm quark mass.
 The results of pseudo-scalar meson dispersion relations at
 different values of $\nu$ then enable us to
 extrapolate/interpolate to the
 optimized value of the bare speed of light parameter $\nu$
 for a given quark mass at a given gauge coupling $\beta$.
 In order to measure the meson
 dispersion relations with better accuracy, asymmetric spatial
 lattice volumes are used which provide us with more
 non-degenerate (in the sense of energy) low-momentum modes.
 The quark action thus obtained
 can then be utilized in future studies on physical
 properties of hadrons with either light or heavy quarks.

 This paper is organized in the following manner. In Section 2,
 a particular form of clover-improved Wilson fermion action on
 anisotropic lattices is introduced. In Section 3,
 the calculation of the energy levels and
 dispersion relations for pseudo-scalar
 meson is discussed. This is performed for quark mass values ranging from
 the strange all the way to the charm quark mass at
 various values of gauge coupling and bare speed of light
 parameter $\nu$.
 By extrapolation or interpolation,
 the optimal values of the bare speed of light
  (denoted by $\nu_{opt}$) can
 then be determined for various values of $\beta$
 for a given bare quark mass parameter.
 In particular, we give the estimates for the
 optimal choice of $\nu$ at the physical charm
 and strange quark mass values for a given $\beta$.
 In Section 4, we will conclude with some general remarks.

\section{Improved Wilson Fermions on Anisotropic Lattices}
\label{sec:fermion}

 Consider a finite four-dimensional lattice with lattice spacing
 $a_\mu$ along the $\mu$ direction with $\mu=0,1,2,3$.
 For definiteness, we denote $a_0=a_t$ and $a_i=a_s$
 for $i=1,2,3$. We will use $\xi = a_s/a_t$ to denote
 the bare aspect ratio of the asymmetric lattice.
 The quark actions on anisotropic lattices have been
 studied extensively in the
 literature~\cite{kronfeld97:aniso,klassen98:wilson_quark,%
 klassen99:aniso_wilson,shigemitsu00:aniso_oneloop,%
 chen01:aniso,kronfeld01:aniso_Oa,umeda01:aniso,%
 chuan01:tune,CPPACS02:aniso,harada02:aniso,okamoto03:aniso,%
 CPPACS03:aniso_unquench,peardon04:aniso}.%
 Using these actions,
 charmed meson spectrum~\cite{mackenzie98:BsDs,juettner03:Ds},
 charmonium spectrum~\cite{chen01:aniso,CPPACS02:aniso},
 charmed baryon spectrum~\cite{lewis01:aniso,nemeto03:aniso_baryon}
 and hadron-hadron scattering
 lengths~\cite{chuan02:pipiI2,chuan04:KN,chuan04:Kpi,chuan04:pipi}
 have been studied.

 We start from the fermion action
 in the hopping parameter parametrization:
 \ba
 \label{eq:Sf_hopping}
 S &=&\bar{\psi}_x M_{xy}\psi_y\;,
 \nonumber \\
 M_{xy}&=&\left[1+\kappa_sc_B\sum_{i<j}\sigma_{ij}\calF_{ij}
 +\kappa_sc_E\sum_{i}\sigma_{0i}\calF_{0i}\right]\delta_{xy}
 \nonumber \\
 &-&\kappa_t\left[(1-\gamma_0)U_0(x)\delta_{x+\hat{0},y}
  +(1+\gamma_0)U^\dagger_0(x-\hat{0})\delta_{x-\hat{0},y}\right]
 \nonumber \\
 &-&\kappa_s\sum_{i}\left[(1-\gamma_i)U_i(x)\delta_{x+\hat{i},y}
  +(1+\gamma_i)U^\dagger_i(x-\hat{i})\delta_{x-\hat{i},y}\right]
  \;.
 \ea
 Here we follow the notation as in Ref.~\cite{kronfeld01:aniso_Oa},
 where we have made the choice $r_t=r_s=1$ for the Wilson
 parameters. Another parameter $\zeta=\kappa_s/\kappa_t$ is
 also commonly used in the literature.
 The forward and backward covariant derivatives
 on the lattice are given by:
 \ba a_\mu \nabla_\mu \psi_x &=& U_\mu (x) \psi_{x+\mu} -\psi_x \;\;,
 \nonumber \\
 a_\mu \nabla^{*}_\mu \psi_x &=& \psi_x - U^\dagger_\mu(x-\mu)
 \psi_{x-\mu} \;\;.
 \ea
 Using these definitions, one can rewrite the fermion
 action~(\ref{eq:Sf_hopping}) in continuum-like notations:
 \ba
 \label{eq:Sf_massterm}
 S &=&\sum_{xy}(a_ta^3_s)
 \bar{\psi}^{(c)}_x M^{(c)}_{xy}\psi^{(c)}_y\;,
 \nonumber \\
 M^{c}_{xy} &\equiv & {M_{xy}\over 2\kappa_ta_t}=
 \left[m_0+{\zeta c_B\over 2a_t}\sum_{i<j}\sigma_{ij}\calF_{ij}
 +{\zeta c_E\over 2a_t}\sum_{i}\sigma_{0i}\calF_{0i}\right]\delta_{xy}
 \nonumber \\
 &+&\gamma_0\left({\nabla_0+\nabla^*_0\over 2}\right)_{xy}
 -{a_t\over 2}(\nabla_0\nabla^*_0)_{xy}
 \nonumber \\
 &+&\sum_i\gamma_i\left({\nabla_i+\nabla^*_i\over 2}\right)_{xy}
 -{\xi\zeta a_s\over 2}(\nabla_i\nabla^*_i)_{xy}
  \;,
 \ea
 where the the continuum fields and the bare quark
 mass $m_0$ are given by:
 \ba
 \bar{\psi}_x &=& a^{3/2}_s{\bar{\psi}^{(c)}_x\over\sqrt{2\kappa_t}}
 \;,\;\; \psi_x=a^{3/2}_s{\psi^{(c)}_x\over\sqrt{2\kappa_t}}\;,
 \nonumber \\
 m_0a_t &=& {1\over 2\kappa_t}-1-3\zeta \;.
 \ea
 For later convenience, we introduce the notation:
 \be
 \label{eq:kcr}
 \nu = \xi\zeta \;,\;\;
 {1\over 2\kappa}= {\xi \over 2\kappa_t}
 =m_0a_s+\xi+3\nu\;.
 \ee
 We call the parameter $\nu$ the bare speed of light parameter.
 The tuning of this parameter will be discussed in the remaining
 part of this paper using pseudo-scalar meson dispersion relations.
 Note that the critical bare quark parameter depends explicitly
 on the parameter $\nu$ even in the free case. This dependence
 also shows up qualitatively in our simulation.

 In quenched calculations, one usually needs to
 calculate the quark propagators at various
 valance quark masses. This amounts to different values of $m_0$ or
 $\kappa$ for the same gauge field configuration. In this case, it
 is convenient to use the following fermion matrix:
 \ba
 \label{eq:shift_M}
 \calM_{xy} &=&\delta_{xy}\sigma + {\mathcal A}_{xy}
 \nonumber \\
 {\mathcal A}_{xy} &=&\delta_{xy}\left[1/(2\kappa_{max})
 +\rho_t \sum^3_{i=1} \sigma_{0i} {\mathcal F}_{0i}
 +\rho_s (\sigma_{12}{\mathcal F}_{12} +\sigma_{23}{\mathcal F}_{23}
 +\sigma_{31}{\mathcal F}_{31})\right]
 \nonumber \\
 &-&\sum_{\mu} \eta_{\mu} \left[
 (1-\gamma_\mu) U_\mu(x) \delta_{x+\mu,y}
 +(1+\gamma_\mu) U^\dagger_\mu(x-\mu) \delta_{x-\mu,y}\right] \;\;,
 \ea
 where the coefficients are given by:
 \ba
 \eta_i &=&{\nu \over 2} \;\;, \;\;
 \eta_0={\xi \over 2} \;\;, \;\;
 \sigma={1 \over 2\kappa}-{1\over 2\kappa_{max}}\;\;,
 \nonumber \\
 \rho_t &=& \nu {(1+\xi) \over 4} \;\;, \;\;
 \rho_s = {\nu \over 2} \;\;.
 \ea
 Here we have used the tree-level, zero quark mass
 relation:~\cite{kronfeld01:aniso_Oa}
 \be
 c_B=1 \;,\;\;
 c_E={1+\xi\over 2}\;,
 \ee
 Note that, in principle the parameters $c_B$
 and $c_E$ also have complicated dependence on the
 bare quark mass which we neglect in this study.
 In this notation, the bare quark mass
 dependence is singled out into parameter $\sigma$ and the
 matrix ${\mathcal A}$ remains {\em unchanged}
 when the bare quark mass is varied.  Therefore, one could utilize the
 shifted structure of the matrix ${\mathcal M}$ to solve for
 quark propagators at various
 values of $m_0$ (or equivalently $\kappa$)
 at the cost of solving only one value of $\kappa=\kappa_{max}$,
 using the so-called  Multi-Mass Minimal
 Residual ($M^3R$ for short) algorithm %
 \cite{frommer95:multimass,glaessner96:multimass,beat96:multimass}.

 To implement the tadpole improvement,
 one replaces each spatial link
 $U_i(x)$ by $U_i(x)/u_s$ while keeping the temporal link unchanged.
 \footnote{One can also introduce the tadpole improvement
 parameter $u_t$ for the temporal link. For large anisotropy
 $\xi$, this turns out to be irrelevant since the temporal
 lattice spacing is small enough and for all practical purposes,
 one can set $u_t=1$.}
 This results in the same fermion matrix~(\ref{eq:shift_M}) except
 that the parameters are replaced by:
 \ba
 \label{eq:parameters_TI}
 \eta_i &=&{\nu \over 2u_s} \;\;, \;\;
 \eta_0={\xi\over 2} \;\;,
 \;\;\sigma={1 \over 2\kappa}-{1\over 2\kappa_{max}}\;\;,
 \nonumber \\
 \rho_t &=& \nu{(1+\xi)\over 4u^2_s} \;\;, \;\;
 \rho_s = {\nu \over 2u^4_s} \;\;.
 \ea
 It is the quark action with these parameters that
 will be studied in this paper numerically.

 \section{Simulation Results}

 In this section, we present our numerical results for
 the study of the pseudo-scalar meson dispersion relations
 for various gauge coupling $\beta$.
 Our main focus lies upon the tuning of
 the bare speed of light parameter $\nu$ for a given
 gauge coupling $\beta$ and a given bare quark mass.
 The parameter $\nu$ has to be tuned such that the lattice
 energy-momentum dispersion relations of pseudo-scalar
 mesons under investigation reproduce the continuum form
 in the low-momentum limit. To achieve this goal, one has
 to go through several procedures which we will describe
 in the following.

 \subsection{Simulation parameters and meson correlation functions}

 The basic parameters of our simulation are summarized in
 Table~\ref{tab:basics}.
 \begin{table}[htb]
 \caption{Simulation parameters for lattices (all with $\xi=5$)
 studied in this work. All lattices are of the
 size $6\times9\times 12\times 50$ except for
 $\beta=3.0$ for which the lattice sizes
 are $8\times 12\times 16\times 50$.
 \label{tab:basics}}
 \begin{center}
 \begin{tabular}{|c|c|c|c|c|c|c|c|c|c|}
  \hline
  $\beta$ & \multicolumn{3}{c|}{$2.2$}
          & \multicolumn{3}{c|}{$2.4$}
          & \multicolumn{3}{c|}{$2.6$} \\
          \cline{2-10}
          & \multicolumn{3}{c|}{$r_0/a_s=1.76$}
          & \multicolumn{3}{c|}{$r_0/a_s=2.18$}
          & \multicolumn{3}{c|}{$r_0/a_s=2.48$} \\
          \hline
  $\nu$   & $0.86$ & $0.90$ & $0.94$
          & $0.85$ & $0.90$ & $0.94$
          & $0.87$ & $0.90$ & $0.93$ \\
          \hline
 $m_{cr}(\nu)$ & $7.723(8)$ & $7.879(7) $ & $8.035(8)$
          & $7.677(9)$ & $7.863(8) $ & $8.021(5)$
          & $7.764(4)$ & $7.878(5) $ & $7.982(6)$ \\
          \hline
 $\kappa$ & $0.0630$ & $0.0620$ & $0.0605$
          & $0.0630$ & $0. 0620$ & $0.0610$
          & $0.0630$ & $0.0620$ & $0.0615$ \\
          &$0.0625$ & $0.0615$ & $0.0600$
          & $0.0625$ & $0.0615$ & $0.0605$
          & $0.0625$ & $0.0615$ & $0.0610$ \\
          &$0.0620$ & $0.0610$ & $0.0595$
          & $0.0620$ & $0.0610$ & $0.0600$
          & $0.0620$ & $0.0610$ & $0.0605$ \\
          & $0.0615$ & $0.0605$ & $0.0590$
          & $0.0615$ & $0.0605$ & $0.0595$
          & $0.0615$ & $0.0605$ & $0.0600$ \\
          &$0.0605$ & $0.0595$ & $0.0580$
          & $0.0605$ & $0.0595$ & $0.0585$
          & $0.0605$ & $0.0595$ & $0.0590$ \\
          &$0.0590$ & $0.0580$ & $0.0565$
          & $0.0595$ & $0.0585$ & $0.0575$
          & $0.0590$ & $0.0585$ & $0.0580$ \\
          & $0.0575$ & $0.0565$ & $0.0550$
          & $0.0585$ & $0.0575$ & $0.0565$
          & $0.0575$ & $0.0570$ & $0.0565$ \\
          &$0.0560$ & $0.0550$ & $0.0535$
          & $0.0575$ & $0.0565$ & $0.0555$
          & $0.0565$ & $0.0555$ & $0.0555$ \\
          &$0.0545$ & $0.0540$ & $0.0525$
          & $0.0565$ & $0.0555$ & $0.0545$
          & $0.0555$ & $0.0545$ & $0.0545$ \\
          & $0.0540$ & $0.0530$ & $0.0520$
          & $0.0560$ & $0.0550$ & $0.0540$
          & $0.0550$ & $0.0540$ & $0.0540$ \\
          &$0.0535$ & $0.0525$ & $0.0515$
          & $0.0555$ & $0.0545$ & $0.0535$
          & $0.0545$ & $0.0535$ & $0.0535$ \\
          &$0.0530$ & $0.0520$ & $0.0510$
          & $0.0550$ & $0.0540$ & $0.0530$
          & $0.0540$ & $0.0530$ & $0.0530$ \\
          \hline
   $\beta$ & \multicolumn{3}{c|}{$2.8 (6\cdot 9\cdot 12\cdot 50)$}
          & \multicolumn{3}{c|}{$2.8 (8\cdot 12\cdot 16\cdot 50)$}
          & \multicolumn{3}{c|}{$3.0$} \\
          \cline{2-10}
          & \multicolumn{3}{c|}{$r_0/a_s=3.13$}
          & \multicolumn{3}{c|}{$r_0/a_s=3.13$}
          & \multicolumn{3}{c|}{$r_0/a_s=4.13$} \\
          \hline
  $\nu$   & $0.87$ & $0.90$ & $0.93$
          & $0.87$ & $0.90$ & $0.93$
          & $0.87$ & $0.90$ & $0.93$\\
          \hline
  $m_{cr}(\nu)$ & $7.744(9)$ & $7.859(6) $ & $7.972(7)$
          & $7.739(6)$ & $7.862(5) $ & $7.972(6)$
          & $7.740(3)$ & $7.850(4) $ & $7.968(4)$\\
          \hline
 $\kappa$ & $0.0635$ & $0.0625$ & $0.0615$
          & $0.0635$ & $0.0625$ & $0.0615$
          & $0.0635$ & $0.0625$ & $0.0615$ \\
          &$0.0630$ & $0.0620$ & $0.0610$
          & $0.0630$ & $0.0620$ & $0.0610$
          & $0.0630$ & $0.0620$ & $0.0610$ \\
          &$0.0625$ & $0.0615$ & $0.0605$
          & $0.0625$ & $0.0615$ & $0.0605$
          & $0.0625$ & $0.0615$ & $0.0605$\\
          & $0.0620$ & $0.0610$ & $0.0600$
          & $0.0620$ & $0.0610$ & $0.0600$
          & $0.0620$ & $0.0610$ & $0.0600$\\
          & $0.0615$ & $0.0600$ & $0.0590$
          & $0.0600$ & $0.0590$ & $0.0580$
          & $0.0600$ & $0.0590$ & $0.0580$\\
          & $0.0610$ & $0.0590$ & $0.0585$
          & $0.0580$ & $0.0570$ & $0.0560$
          & $0.0580$ & $0.0570$ & $0.0560$\\
          & $0.0600$ & $0.0580$ & $0.0575$
          & $0.0575$ & $0.0565$ & $0.0555$
          & $0.0575$ & $0.0565$ & $0.0555$\\
          & $0.0590$ & $0.0570$ & $0.0565$
          & $0.0570$ & $0.0560$ & $0.0550$
          & $0.0570$ & $0.0560$ & $0.0550$\\
          & $0.0580$ & $0.0565$ & $0.0555$
          &  &   &
          &  &  & \\
          & $0.0570$ & $0.0560$ & $0.0550$
          &   &   &
          &  &  & \\
          & $0.0565$ & $0.0555$ & $0.0545$
          &   &   &
          &  &  & \\
          & $0.0560$ & $0.0550$ & $0.0540$
          &  &   &
          &  &  & \\
           \hline
 \end{tabular}
 \end{center}
 \end{table}
 For the study of pseudo-scalar meson dispersion relations,
 it is advantageous to use lattices with asymmetric
 three-volume. This provides more non-degenerate
 (in the sense of its energy)
 low-momentum modes than the conventional symmetric volumes.
 All lattices in this study
 are of the size $6\times 9\times 12\times 50$ except for
 the lattices at $\beta=3.0$ where $8\cdot 12\cdot 16\cdot 50$
 lattices are studied.
 \footnote{In our preliminary studies,
 $4\times 6\times 8\times 40$ have also been simulated.
 We choose to present our results for larger lattices since
 they yield better accuracy for the pseudo-scalar meson
 dispersion relation measurements.}
 To further check finite volume effects,
 a low statistics run (about $120$ configurations)
 for $\beta=2.8$ with
 larger lattice volumes was also performed.
 It turns out that the light meson mass values
 are somewhat modified but the final result
 of the optimal value of $\nu$ remain compatible
 within errors (see Table~\ref{tab:mass_parameters}).
 The aspect ratio is
 $\xi=a_s/a_t\simeq\xi_0=5$ for all lattices.
 The value of $\beta$ ranges between $2.2$ and $3.0$,
 roughly corresponding to spatial lattice spacing $a_s$
 between $0.12$ and $0.27$fm in physical units.
 For each particular value of $\beta$,
 gauge field configurations are generated using the conventional
 pseudo-heatbath algorithms with over-relaxation.

 For gauge field configurations at
 a given value of $\beta$, $3$ different values of
 the bare speed of light parameter $\nu$ as shown in
 Table~\ref{tab:basics} are studied.
 Quark propagators with zero and non-zero three-momenta
 are obtained using the $M^3R$ algorithm
 with wall sources for each dirac and color index.
 With the help of the $M^3R$ algorithm, by solving only
 one linear equation, one obtains
 the quark propagators with $12$ (in the case of $\beta=3.0$,
 only $8$ values of $\kappa$ were taken) different values of
 $\kappa$ which correspond to different quark masses.
 The values of $\kappa$ are chosen such that the
 quark mass values range roughly from around the
 physical strange quark mass all the way up to
 the physical charm quark mass. The values of $\kappa$
 for each parameter set $(\beta,\nu)$ are
 also tabulated in Table~\ref{tab:basics}.

 In this paper, we focus on the single pseudo-scalar
 states with definite three-momentum.
 We define the pseudo-scalar and vector meson operators
 as follows:
 \be
 P(\bx,t)=\bar{q}_1(\bx,t)\gamma_5q_2(\bx,t)\;,
 \ee
 where $q_1$, $q_2$ ($\bar{q}_1$, $\bar{q}_2$)are
 quark field operators of two (possibly identical) flavors.
 Operators which create meson states with definite
 three-momentum $\bk$ are then defined accordingly:
 \be
 P(\bk,t)={1\over \sqrt{V_3}}\sum_\bx
 P(\bx,t)e^{-i\bk\cdot\bx}\;,\;\;
 \ee
 where $V_3$ designates the three-volume of the lattice.
 Using these operators, one constructs the corresponding
 meson correlation function:
 \be
 C^{(PS)}(\bk,t)=\langle P(\bk,t)^{\dagger}P(\bk,0)\rangle\;,
 \ee
 Using Wick's theorem, the above defined correlation function
 can be expressed in terms of the quark propagators:
 \be
 C^{(PS)}(\bk,t)= {1\over V_3}\sum_\by
 Y^{(1)\rho b}_{\beta a\by t}
 \left[X^{(2)\sigma b}_{\alpha a\by t}\right]^*
 \cdot e^{i\bk\cdot\by}\;,
 \ee
 where the Greek subscripts/superscripts in
 the solution vectors $X$ and $Y$ are
 Dirac indices while Roman subscripts/superscripts
 are color indices.
 The solution vectors $X$ and $Y$ are given by
 the inverse fermion matrix elements:
 \ba
 Y^{\rho b}_{\beta a\by t} &=&
 \sum_\bx \calM^{-1}_{\beta a\by t;\rho b\bx 0}
 e^{-i\bk\cdot\bx}\;,
 \nonumber \\
 X^{\rho b}_{\beta a\by t} &=&
 \sum_\bx \calM^{-1}_{\beta a\by t;\rho b\bx 0}
 \;.
 \ea
 The superscript $(1)$ or $(2)$ on these solution
 vectors indicates that the quark mass should
 be that of quark flavor $1$ or $2$ (possibly the same).
 These solution vectors are obtained by solving the linear
 equation of the fermion matrix $\calM$ with a suitable wall source.

 \begin{figure}[tb]
 \begin{center}
 \includegraphics[height=16.0cm,angle=0]{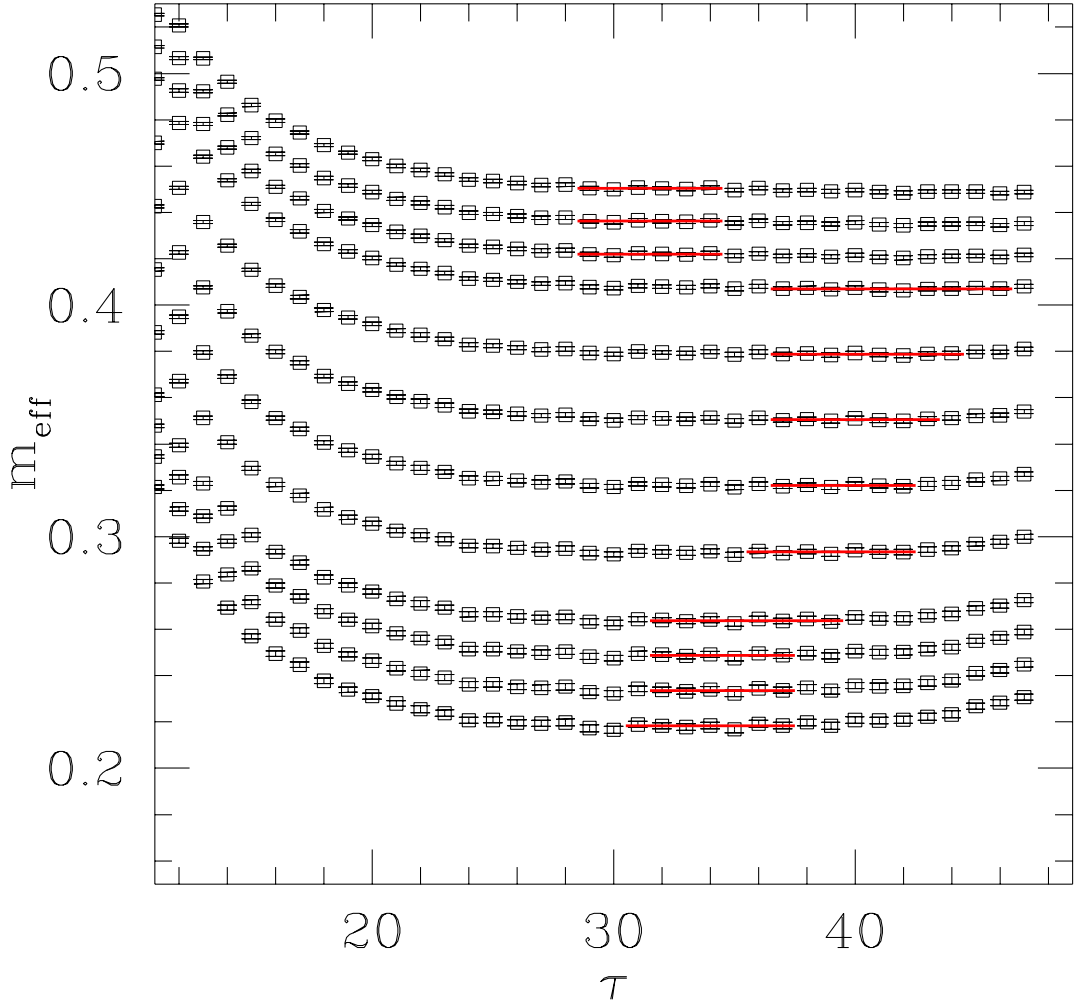}
 \end{center}
 \caption{ The effective mass plots for the
 pseudo-scalar meson made up of a quark and
 an anti-quark for zero three-momenta at
 $\beta=2.4$, $\nu=0.85$.
 With one quark mass parameter $m_1$ being fixed,
 different lines in the plot correspond to
 different values of the other quark mass $m_2$.
 The red horizontal bars in the plot indicates the ranges
 in which the mass of the meson are extracted.}
 \label{fig:ekfit.plot}
 \end{figure}
 The energy values of a pseudo-scalar meson
 with definite three-momentum $\bk$
 (including zero-momentum) is
 obtained from their respective correlation
 functions $C^{(PS)}(\bk,t)$ by
 finding the plateaus in their effective mass plots.
 In Fig.~\ref{fig:ekfit.plot}, we show the effective mass plots
 of a pseudo-scalar meson with zero three-momentum for
 $\beta=2.4$, $\nu=0.85$.
 The pseudo-scalar meson consists of a quark and an anti-quark with
 all possible bare quark mass value combinations $(m_1,m_2)$.
 In Fig~~\ref{fig:ekfit.plot}, we illustrate one of the
 situations with $m_1$ being fixed. The effective mass
 plots are shown for all $m_2$ values.
 Different lines in the plot then
 correspond to different values of the other bare
 quark mass parameter $m_2$.
 There are $12$ lines in each of these windows
 which correspond to $12$
 different values of the bare quark mass $m_2$ being simulated.
 It is seen that all effective mass plots develop
 nice plateaus at large
 temporal separation and accurate values of the
 pseudo-scalar energy $E_{PS}(m_1,m_2,\bk)$ ( and also the mass
 $M_{PS}(m_1,m_2)=E_{PS}(m_1,m_2,\bk=0)$ ) can thus be extracted.
 The red horizontal bars in the plot indicate the ranges
 in which the meson energy values are extracted.
 The errors for the data points in this plot are analyzed
 using the standard jack-knife method. The intervals from which
 we extract the energy values are self-adjusted according
 to the minimum of $\chi^2$ per degree of freedom.
 The quality of the effective mass plots
 for other parameter sets are similar.

 \subsection{Obtaining the pseudo-scalar meson energy at fixed quark masses}

 From the effective mass plots of pseudo-scalar meson correlation
 functions we obtain the energy values
 of a single pseudo-scalar meson with
 definite three-momentum $\bk$:
 $E^2_{PS}(m_1,m_2,\nu,\bk)$.
 We thus have these energy values for each $\beta$, $\nu$, $\bk$
 and all possible bare quark mass values $(m_1,m_2)$ which we choose
 to calculate the meson correlation functions.
 Here the bare quark mass parameter is defined via:
 \be
 \label{eq:m_def}
 m\equiv {1\over 2\kappa}-{1\over 2\kappa_{cr}(\nu)}\;.
 \ee
 where $\kappa_{cr}(\nu)$ is the critical hopping parameter
 at which the pion mass vanishes for a particular $(\beta,\nu)$.
 Note that this value depends on $\nu$ for a given $\beta$.
 For a given value of $\beta$,
 the critical hopping parameter
 $1/(2\kappa_{cr}(\nu)$ for each $\nu$ is obtained by
 fitting the pion (made up of equal mass quarks)
 mass squared versus $1/(2\kappa)$ using a quadratic function
 in the low quark mass region. From these fits, we obtain
 the critical value $1/(2\kappa_{cr}(\nu))$ for each $\nu$
 at a given $\beta$.

 The reason that we choose the bare quark mass
 parameter $m$ instead of
 the hopping parameter itself is the following.
 Our goal is to find the optimal values of $\nu$ such
 that the pseudo-scalar meson exhibits the proper dispersion
 relation in the low-momentum region.
 Therefore, for a given value of $\beta$,
 we want to fix the quark mass values and
 interpolate/extrapolate in $\nu$ to obtain the optimal
 value of $\nu$ at which the meson dispersion relation has the right form.
 This should be done for all possible quark mass values.
 It is better to perform this interpolation/extrapolation for
 fixed $m$ instead of fixed hopping parameter
 pair $\kappa$ since the critical hopping parameters
 themselves depend explicitly on $\nu$,
 as is seen evidently from the tree-level relation
 Eq.~(\ref{eq:kcr}). This dependence is also seen from our
 simulation. So for different values of $\nu$,
 the same value of $\kappa$ for different $\nu$ in fact corresponds to
 {\em different} bare quark mass values.
 Therefore, it is more appropriate to interpolate/extrapolate
 in $\nu$ for fixed bare quark mass parameter pair $m$.
 Note that the bare quark mass $m$ as defined
 in~(\ref{eq:m_def}) is an {\em independent} parameter of
 the quark action. In other words, no matter what
 the value of $1/(2\kappa_{cr}(\nu))$ comes out to be
 for each $\nu$, we could choose values of $m$,
 independent of $\nu$, since we are free to adjust the
 set of values for $\kappa$.

 \begin{figure}[tb]
 \begin{center}
 \includegraphics[height=16.0cm,angle=0]{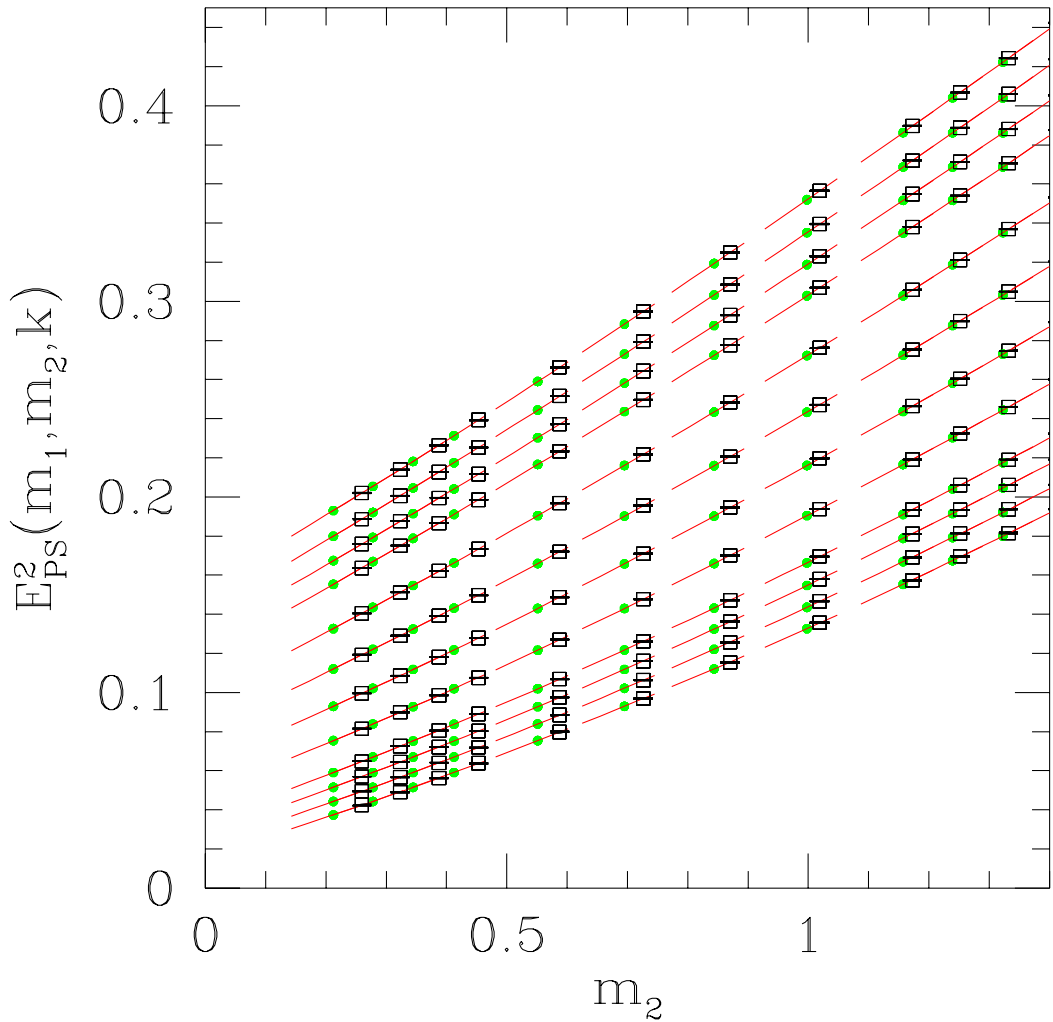}
 \end{center}
 \caption{ Interpolation of the pseudo-scalar meson energy
 squared $E^2_{PS}(m_1,m_2,\nu,\bk)$
 versus the quark mass parameter
 pairs $(m_1,m_2)$ to the common values $(\bar{m}_1,\bar{m}_2)$ is
 shown for $\beta=2.4$, $\nu=0.85$. Only the zero three-momentum
 case is shown in this plot. Values of
 $E^2_{PS}(m_1,m_2,\nu,\bk=0)$ are shown as data points.
 Red lines are the quadratic interpolations around each value
 of $\bar{m}$ using $6$ points of $m$ close to it.}
 \label{fig:showshift}
 \end{figure}
 It turns out that our choices of the hopping
 parameters for different $\nu$ are such that the range of
 $m$ are roughly the same for a given $\beta$
 while the individual values are not identical.
 Therefore, before we make any interpolation/extrapolation in $\nu$,
 which should be done at fixed $m$ for all $\nu$, we
 first have to interpolate the energy values
 $E^2_{PS}(m_1,m_2,\nu,\bk)$ at different $\nu$ to the
 the {\em same} quark mass values: $(\bar{m}_1,\bar{m}_2)$.
 In the analysis, we pick the values of $\bar{m}$ to
 be the average of the three corresponding values
 at three different $\nu$.
 Of course, the choice of the
 common values for $\bar{m}$ at different $\nu$ is somewhat
 arbitrary and any other choice is equally well as
 long as the range of $\bar{m}$ roughly coincides
 with the ranges of the $m$ for different $\nu$
 such that the interpolation can be done reliably.
 The interpolation of the
 energy values squared $E^2_{PS}(m_1,m_2,\nu,\bk)$ is
 performed by a quadratic interpolation in the
 quark mass parameter using 4-6 neighboring points close to
 the values of $\bar{m}$.
 It is checked that all interpolations yield good
 fitting qualities.
 In Fig.~\ref{fig:showshift}, we show this interpolation
 for $\beta=2.4$, $\nu=0.85$. Values of
 $E^2_{PS}(m_1,m_2,\nu,\bk)$
 are shown as data points in the plot which
 correspond to zero three-momentum.
 The green points are the values of $\bar{m}$ to which
 $E^2_{PS}(m_1,m_2,\nu,\bk)$ are interpolated.
 Red line segments are the quadratic interpolation in the
 quark mass parameter using $4-6$ points around each
 value of $\bar{m}$. The number of points being taken
 in each interpolation is determined by the condition of
 minimum $\chi^2$ per degree of freedom.
 The situation for non-zero momentum is similar.
 All interpolation yields good fitting quality.

 As the outcome of this procedure,
 we have all the quantities:
 $E^2_{PS}(\bar{m}_1,\bar{m}_2,\nu,\bk)$ that are
 at the same sequence of points $\bar{m}$ for
 different values of $\nu$.
 This procedure is performed for every value of
 $\beta$ and for every three-momentum mode under
 investigation (see Eq.~(ref{eq:ns}) for the
 momentum modes being studied). In the discussion below, we will drop
 the bars on the quark mass parameters for simplicity
 with the understanding that all energy levels
 $E_{PS}(m_1,m_2,\nu,\bk)$ are already
 interpolated to the same set of $(m_1,m_2)$ for
 different $\nu$.

 \subsection{Extraction of $Z$ parameter}

 In this work, we utilize lattices with asymmetric volumes.
 This asymmetry helps to break the cubic symmetry
 in the momentum space and lifts the degeneracy of the
 meson energies. For example, by using a
 lattice of size $6\times 9\times 12\times 50$, we
 have $8$ non-degenerate low-momentum modes,
 compared with only $4$ with symmetric volumes.
 This technique also proves to be useful for the measurement of
 other momentum-dependent quantities like
 the hadron-hadron scattering phase
 shifts~\cite{chuan04:asymmetric,chuan04:asymmetric_long}.
 The three-momenta $\bk$ in an asymmetric box of size
 $L_1\times L_2\times L_3$
 \footnote{For definiteness, we pick $L_1\leq L_2\leq L_3$.}
 are quantized according to:
 \be
 \label{eq:quantize_k}
 \bk=\left({2\pi\over L_1}n_1,
 {2\pi\over L_2}n_2,{2\pi\over L_3}n_3\right)
 \;,
 \ee
 with $\bn=(n_1,n_2,n_3)\in \mathbb{Z}^3$ being
 three-dimensional integers. In this work,
 the energy values of a meson with the following
 $8$ three-momentum are measured:
 \ba
 \label{eq:ns}
 \bn &=&(0,0,0)\;,(0,0,1)\;,(0,1,0)\;,(1,0,0)\;,
 \nonumber \\
     & &(0,1,1)\;,(1,0,1)\;,(1,1,0)\;,(1,1,1)\;.
 \ea

 For a given three-momentum $\bk$,
 the pseudo-scalar meson energy levels $E^2_{PS}(m_1,m_2,\nu,\bk)$
 are fitted according to the expected continuum dispersion relation:
 \be
 \label{eq:dis}
 E^2_{PS}(m_1,m_2,\nu,\bk) = M^2_{PS}(m_1,m_2,\nu)
 + Z_{PS}(m_1,m_2,\nu) \bk^2\;\;,
 \ee
 in the low-momentum region.
 The fitting is performed for pseudo-scalar mesons with
 all possible bare quark mass combinations: $(m_1,m_2)$.
 As a result of these fits, we obtain all the $Z$ parameters
 of the pseudo-scalar mesons as a function of two
 bare quark mass parameters: $Z_{PS}(m_1,m_2,\nu)$.
 In particular, for the pseudo-scalar meson made up of
 quarks with the same mass, the corresponding $Z$ parameter
 depends on one quark mass parameter, namely $Z_{PS}(m,m,\nu)$.
 \begin{figure}[tb]
 \begin{center}
 \includegraphics[height=16.0cm,angle=0]{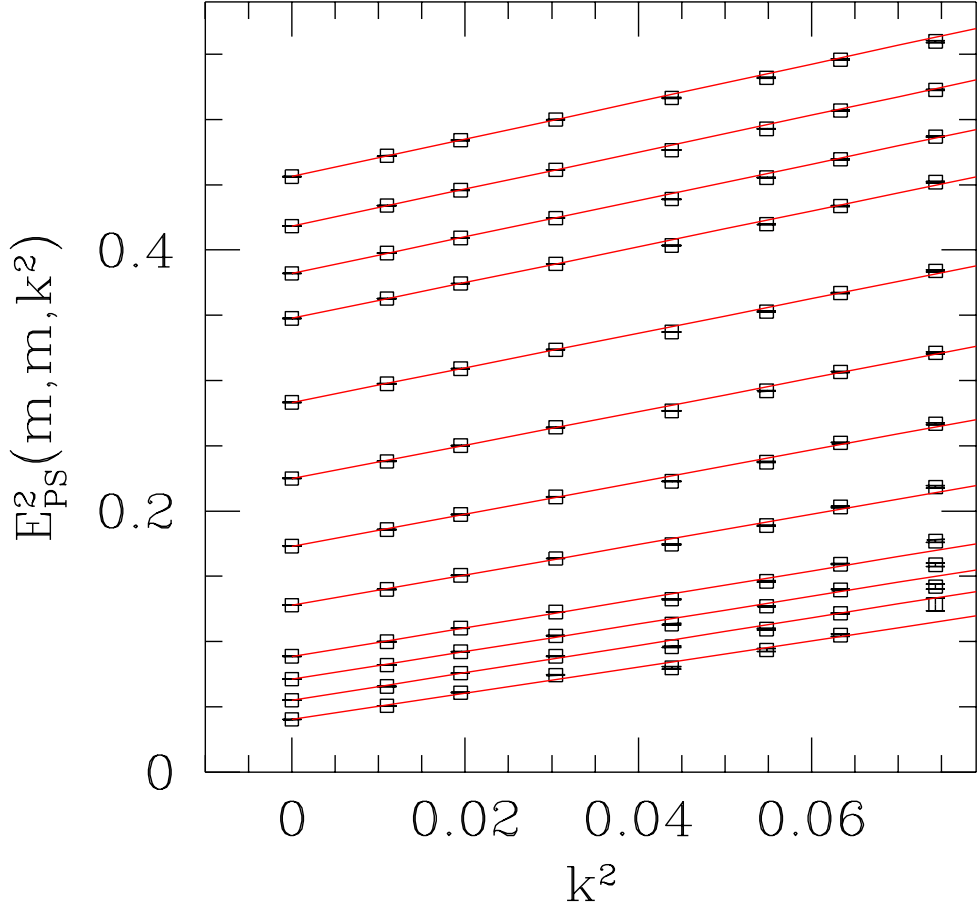}
 \end{center}
 \caption{ Dispersion relations of the pseudo-scalar meson
 at $\beta=2.4$, $0.85$.
 The data points show the pseudo-scalar meson energy squared
 $E^2_{PS}(m,m,\nu,\bk)$ at a given
 three-momentum $\bk$ versus $\bk^2$.
 The straight lines are the linear fits to the data with the fitting
 range starting from the low-momentum end and self-adjusted according
 to the minimum $\chi^2$ per degree of freedom. The slope of the
 lines then yield the desired $Z$ parameters.
 Different lines in the plot
 indicate the linear fits for different bare quark mass
 parameter $m$.}
 \label{fig:dis}
 \end{figure}
 In Fig.~\ref{fig:dis},
 the linear fits of dispersion relations $E^2_{PS}(m,m,\nu,\bk^2)$ versus
 $\bk^2$ for the pseudo-scalar
 meson with equal mass quark and anti-quark
 are shown for $\beta=2.4$, $\nu=0.85$.
 The straight lines in the plot represent the linear
 fits for $12$ values of bare quark parameter $m$.
 The linear fits utilize
 the low-momentum data points (including the zero-momentum point)
 according to Eq.~(\ref{eq:dis}) and the fitting range
 for each line is self-adjusted to yield the minimum $\chi^2$ for each
 degree of freedom. The slope of these lines then
 yield the parameters $Z_{PS}(m,m,\nu)$ for all bare quark
 mass parameter $m$. Fitting qualities for other $(\beta,\nu)$
 are quite similar. We have also tried another (conventional)
 way of extracting the $Z$ parameters,
 namely by using only the zero-momentum point and
 the lowest non-vanishing momentum point ($\bn=(0,0,1)$ in this case).
 This is what has been done in the
 literature by many authors~\cite{harada02:aniso,CPPACS03:aniso_unquench,%
 okamoto03:aniso,peardon04:aniso}.
 We find that the $Z$ parameters are always
 better determined by using linear fits with more momentum points
 as compared with only two lowest momentum points.
 Therefore, we see the advantage of using asymmetric volumes
 for all our parameter sets.

 \subsection{Finding the optimal values of $\nu$}

 The optimal choice for the bare speed of light
 parameter $\nu$ in the quark action is
 determined from the corresponding
 pseudo-scalar meson dispersion relations,
 or more explicitly, from the $Z$ parameters
 $Z_{PS}(m,m,\nu)$ extracted from dispersion relations
 which is discussed the previous subsection.
 One requires that the dispersion
 relation of the pseudo-scalar meson made up of the same quark
 flavor reproduces its continuum
 counter-part in the low-momentum limit.
 That is to say, the optimal choice of $\nu$ has to
 be such that:
 \be
 Z_{PS}[m,m,\nu_{opt}(m)]=1\;.
 \ee
 This yields the optimal value of $\nu$ as a function
 of the bare quark mass parameter: $\nu_{opt}(m)$.

 In practice, we use the values of $Z_{PS}(m,m,\nu)$ at
 different $\nu$ and perform a linear extrapolation/interpolation
 in $\nu$ for every value of $m$.
 \begin{figure}[tb]
 \begin{center}
 \includegraphics[height=16.0cm,angle=0]{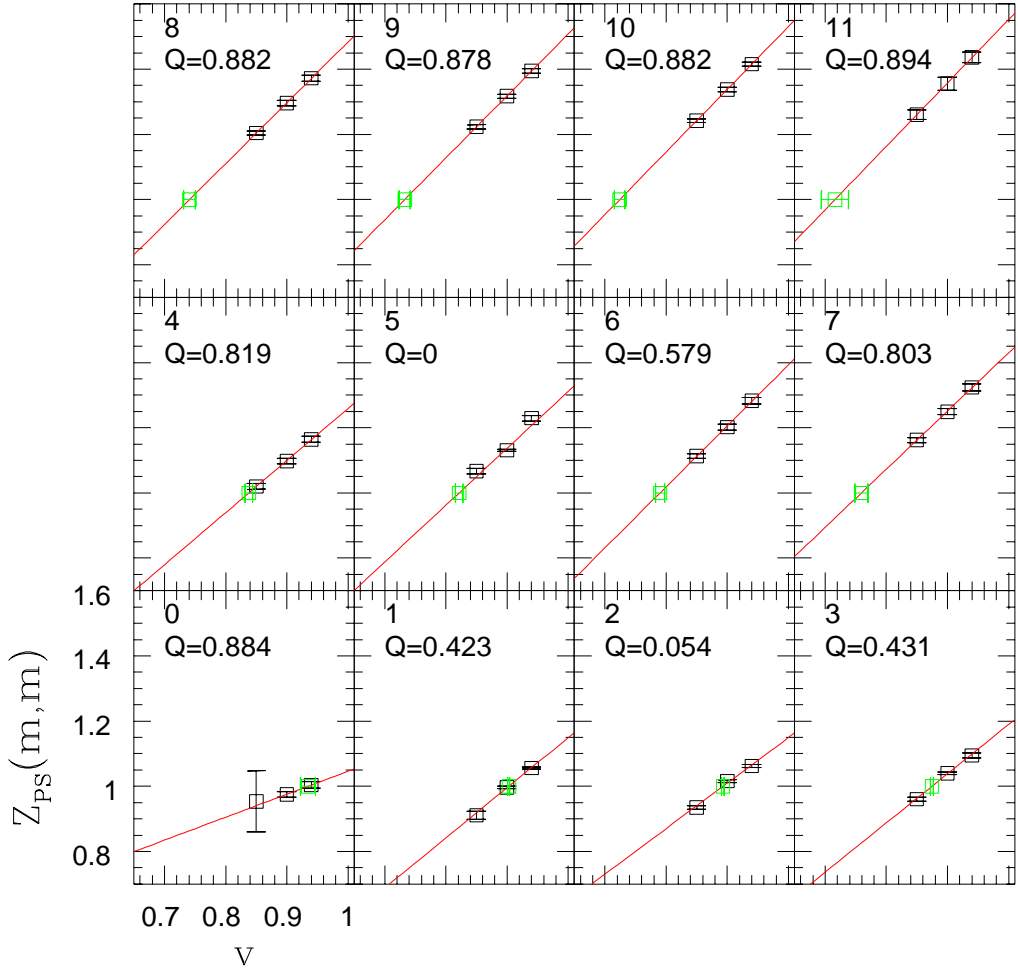}
 \end{center}
 \caption{ Determination of optimal speed of light
 parameter $\nu_{opt}(m)$ is shown for pseudo-scalar
 meson made of equal mass quark and anti-quark. Each small
 window corresponds to different values of bare
 quark mass parameter $m$. Data points in each window are
 the values of $Z_{PS}(m,m,\nu)$. They are fitted
 linearly versus $\nu$ and the optimal values of $\nu$
 are determined by the condition: $Z_{PS}[m,m,\nu_{opt}(m)]=1$
 for each given $m$.
 The results of $\nu_{opt}(m)$ are shown as green
 points in each window together with the corresponding
 error bar. In each window (corresponding to different $m$), the
 quality of the linear fit is also indicated.}
 \label{fig:Znu_equalm}
 \end{figure}
 This is shown in Fig~\ref{fig:Znu_equalm} in the
 case of $\beta=2.4$.
 Each small window in this plot corresponds to different values of bare
 quark mass parameter $m$. Data points in each window are
 the values of $Z_{PS}(m,m,\nu)$ obtained from the pseudo-scalar
 dispersion relations. They are fitted
 linearly versus $\nu$ and the optimal values of $\nu$
 are determined by the condition: $Z_{PS}[m,m,\nu_{opt}(m)]=1$.
 The results of $\nu_{opt}(m)$ for various $m$ are shown as green
 points together with the corresponding error bars.
 In each window (corresponding to different $m$), the
 quality of the linear fit is also shown.

 \subsection{Finding the physical bare quark mass parameters}

 \begin{table}[htb]
 \caption{Extracted physical bare quark mass parameters
 for the charm and the strange quark and the
 optimal values for the bare speed of light parameter
 $\nu$ at physical quark masses for different $\beta$.
 Two sets of data (I) and (II) for $\beta=2.8$ corresponds to
 smaller ($6\cdot 9\cdot 12\cdot 50$) and larger
  ($8\cdot 12\cdot 16\cdot 50$)
 lattices, respectively.
 \label{tab:mass_parameters}}
 \begin{center}
 \begin{tabular}{|c|c|c|c|c|c|c|}
 \hline
 $\beta$                  & $2.2$      & $2.4$      & $2.6$      & $2.8(I)$   & $2.8(II)$      & $3.0$ \\
 \hline\hline
 $m^{(phy)}_c$            & $1.879(2)$ & $1.470(2)$ & $1.194(1)$ & $0.969(1)$ & $0.978(5)$ & $0.677(1)$\\
 \hline
 $m^{(phy)}_s$            & $0.217(3)$ & $0.163(1)$ & $0.155(1)$ & $0.120(2)$ & $0.15(2)$ & $0.073(2)$\\
 \hline
 $\nu_{opt}(m^{(phy)}_c)$ & $0.68(3)$  & $0.71(3)$  & $0.754(5)$ & $0.75(1)$  & $0.74(3)$ & $0.84(1)$\\
 \hline
 $\nu_{opt}(m^{(phy)}_s)$ & $0.940(4)$ & $0.933(3)$ & $0.928(3)$ & $0.98(2)$  & $0.91(4)$ & $0.99(1)$\\
 \hline
 \end{tabular}
 \end{center}
 \end{table}

 For phenomenological reasons, one is particularly interested
 in the optimal parameters of $\nu$ near bare quark mass
 values that correspond to the physical interesting cases.
 In particular, we are interested in the
 values of $\nu_{opt}(m)$ at physical bare strange quark and
 bare charm quark mass,
 namely at $m=m^{(phy)}_s$ or $m=m^{(phy)}_c$.
 To find out this correspondence, one has to investigate the
 bare quark mass dependence of the meson mass
 and fix the bare quark mass parameter $m$ which corresponds to
 the physical case.

 \begin{figure}[tb]
 \begin{center}
 \includegraphics[height=16.0cm,angle=0]{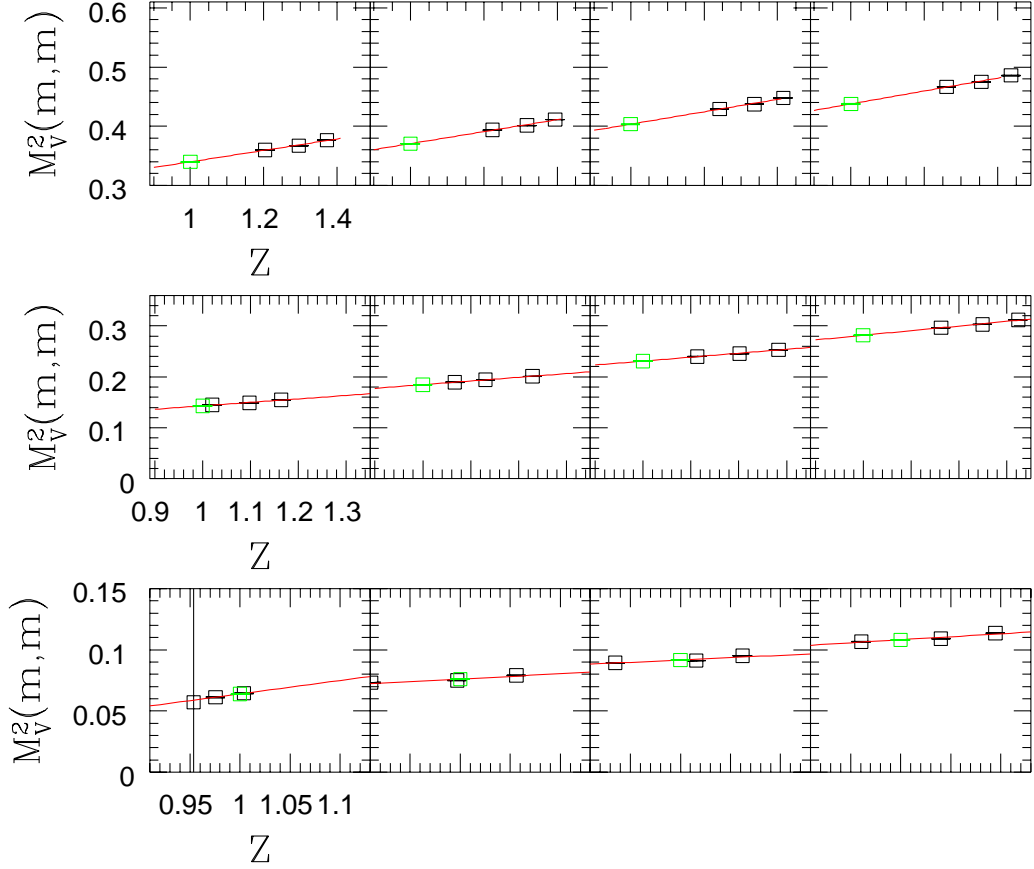}
 \end{center}
 \caption{ Linear extrapolations of
 $M^2_{V}(m,m,\nu)$ versus $Z_{PS}(m,m,\nu)$
 for each value of $m$ at $\beta=2.4$.
 Different windows corresponds to different values
 of $m$. The linear extrapolations are shown by the
 straight lines in the windows.}
 \label{fig:Mnu}
 \end{figure}
 To fix the physical bare quark mass parameters
 for the charm and the strange quark,
 we use the vector meson mass values.
 From the physical $J/\psi$ mass and the
 physical $\Phi$ meson mass, we can fix
 the bare quark mass parameters for the charm
 and the strange respectively.
 First, the vector meson mass
 squared at the optimal value of $\nu$:
 $M^2_{V}[m,m,\nu_{opt}(m)]$ is obtained by
 extrapolating to the optimal value of $\nu$.
 Once this quantity is at our disposal,
 we can perform extrapolation/interpolation in the bare
 quark mass $m$ to find out the physical bare quark mass
 for the charm and the strange quark.
 We therefore extrapolate $M^2_{V}(m,m,\nu)$ versus
 $Z_{PS}(m,m,\nu)$ linearly for each given value of $m$.
 The situation is shown in Fig~\ref{fig:Mnu} for $\beta=2.4$.
 Different small windows corresponds to different values
 of $m$. The linear extrapolations are shown by the
 straight lines in the windows. The extrapolated values of
 $M^2_{V}(m,m,\nu)$ at $Z_{PS}(m,m,\nu)=1$,
 then give the quantities $M^2_{V}[m,m,\nu_{opt}(m)]$.
 The results of $M^2_{V}[m,m,\nu_{opt}(m)]$ for all $m$ are
 then utilized in the quark mass interpolation/extrapolation.

\begin{figure}[tb]
 \begin{center}
 \includegraphics[height=16.0cm,angle=0]{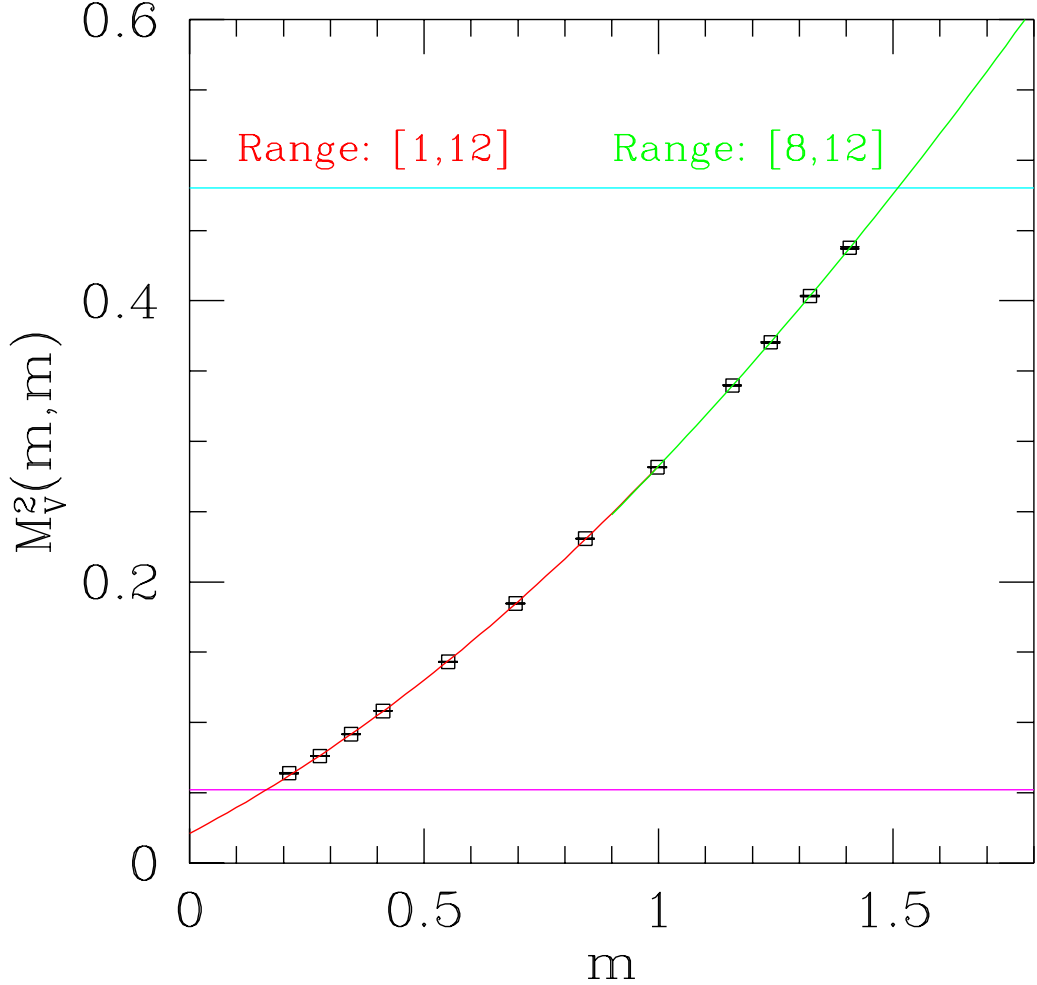}
 \end{center}
 \caption{ Quadratic extrapolations of
 $M^2_{V}[m,m,\nu_{opt}(m)]$ versus
 the bare quark mass $m$ at $\beta=2.4$.
 The red line indicates a fit in the lower quark mass
 region. The green line is the corresponding fit
 in the heavy quark mass region. The corresponding fitting
 ranges are also shown in the figure.
 The fitting ranges are self-adjusted to yield minimum
 $\chi^2$ per degree of freedom.
 The blue horizontal line indicates the value for the
 physical $J/\psi$ meson while the pink horizontal
 line corresponds to the value of the physical $\Phi$ meson.
 The intersect with the green line and the red
 line then yields the estimate for $m^{(phy)}_c$
 and $m^{(phy)}_s$, respectively.}
 \label{fig:Fixqq}
 \end{figure}

 In this work, we perform two quadratic fits for the meson mass
 versus the bare quark mass parameter $m$,
 one in the low quark mass region, the other
 in the heavy quark mass region. We always take the
 fitting form to be:
 \be
 \label{eq:chiral_pion}
 a^2_tM^2_{PS/V}[m,m,\nu_{opt}(m)]= A+Bm+Cm^2\;.
 \ee
 In all our cases, we find the fit parameter $A$ for
 the pseudo-scalar meson in the low quark mass region is
 always consistent with zero as it should be.
 In Fig.~\ref{fig:Fixqq}, we show this extrapolation
 for the vector meson mass at $\beta=2.4$.
 The red line in the plot indicates a fit in the lower quark mass
 region. The green line is the corresponding fit
 in the heavy quark mass region. The corresponding fitting ranges
 are also shown in the figure.
 The fitting ranges are self-adjusted to yield minimum
 $\chi^2$ per degree of freedom.
 To obtain the physical charm quark mass parameter $m^{(phy)}_c$,
 we draw a horizontal line in this figure at the physical
 $J/\psi$ mass: $a^2_tM^2_{J/\psi}$.
 This is obtained by setting the scale using some physical quantity.
 In this work, we choose the hadronic scale $r_0=0.5$fm (the so-called
 Sommer scale) to set the physical scale.
 \footnote{We have assumed that $r_0=0.5$fm exactly. Therefore,
 errors in this scale are not taken into account in the following
 error analysis.}
 For different values of gauge coupling $\beta$,
 the values of $r_0/a_s$ are known from the
 literature~\cite{colin99,chuan01:gluea}
 which are also listed in Table~\ref{tab:basics} for reference.
 With this information, we know the physical meson masses
 in lattice unit. The blue horizontal line in the figure
 representing the value for physical $J/\psi$ intersects
 with the green line and the intersection point
 then yields the estimate for the physical
 charm quark mass parameter $m^{(phy)}_c$.
 Similarly, the pink horizontal line is at the
 value of physical $\Phi$ meson and the intersection
 point with the red line in the lower quark mass region
 yields the estimate for $m^{(phy)}_s$.
 The value of $m^{(phy)}_c$ and $m^{(phy)}_s$
 thus obtained are listed
 in Table~\ref{tab:mass_parameters} for all $\beta$.

 \subsection{Optimal values of $\nu$ at physical quark mass parameters}

 After fixing the physical bare quark mass parameters
 for both the charm and the strange, we can obtain the
 optimal values of the speed of light parameter for
 these cases. In our notation, they correspond to
 $\nu_{opt}(m^{(phy)}_c)$ and
 $\nu_{opt}(m^{(phy)}_s)$, respectively.
 To get these values, we make interpolation/extrapolations
 of $\nu_{opt}(m)$ versus the bare quark mass
 parameter $m$ using quadratic functions in these parameters.
 We choose the appropriate range (heavy quark mass region and
 light quark mass region) for different cases.

\begin{figure}[tb]
 \begin{center}
 \includegraphics[height=16.0cm,angle=0]{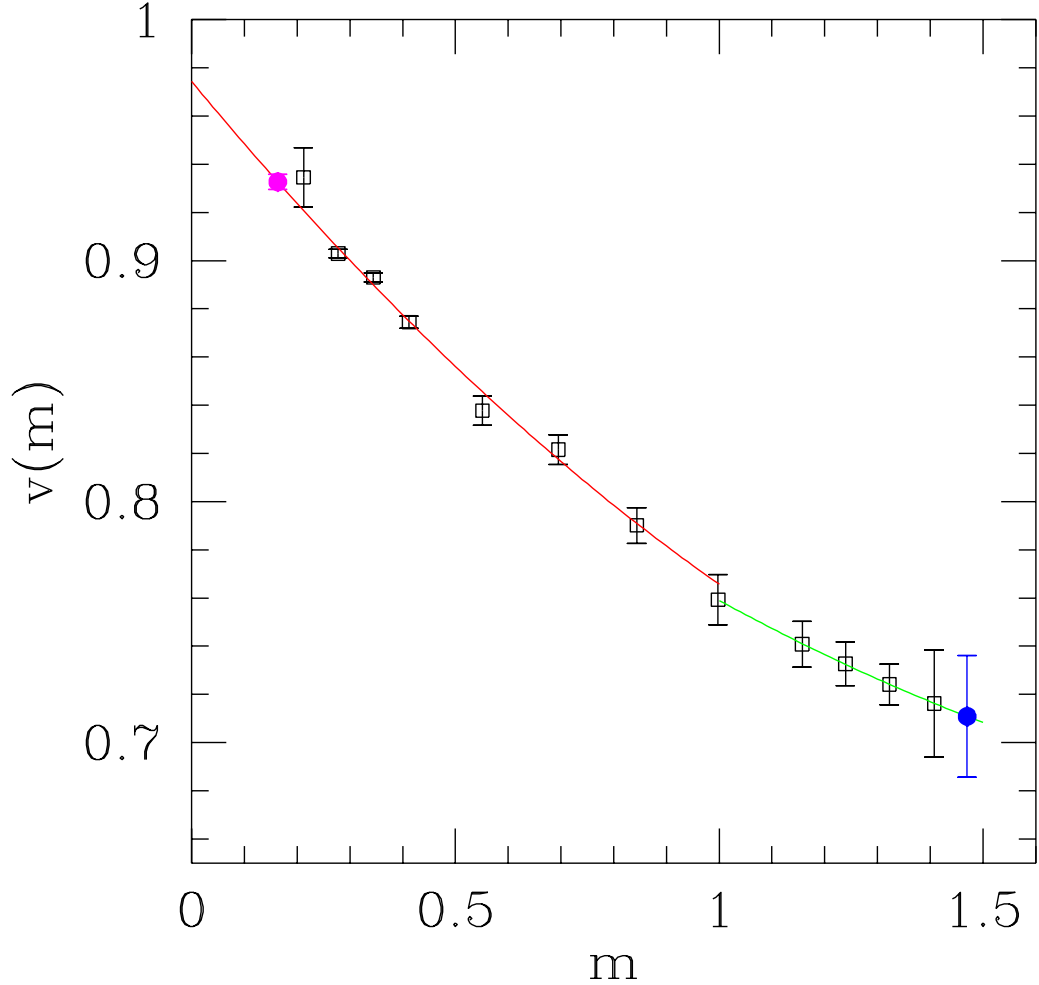}
 \end{center}
 \caption{ The result of $\nu_{opt}(m)$ is shown
 versus the bare quark mass $m$ for $\beta=2.4$.
 The red and the green line is a quadratic fit to the data points
 in the light and the heavy quark mass region respectively.
 At the position of
 the physical bare charm quark mass $m^{(phy)}_c$ and
 the physical bare strange quark mass $m^{(phy)}_s$, the
 result of $\nu_{opt}(m^{(phy)}_c)$ and
 $\nu_{opt}(m^{(phy)}_s)$ are shown
 by the blue and the pink solid point with errors, respectively.}
 \label{fig:Nucc}
 \end{figure}

 In Fig~\ref{fig:Nucc} we show the extrapolation
 of $\nu_{opt}(m)$ versus $m$ in both the light and
 the heavy quark mass region at $\beta=2.4$.
 The red line is the quadratic fit to
 the data points in the low quark mass region while the
 green line is the fit in the heavy quark mass region.
 The final result of
 $\nu_{opt}(m^{(phy)}_c)$ and
 $\nu_{opt}(m^{(phy)}_s)$ are indicated by the
 blue and the pink solid point with errors at the position of
 the physical charm and strange quark mass respectively.
 The results for other values of $\beta$ are summarized in
 Table~\ref{tab:mass_parameters}.

 For the $\beta=2.8$ lattices, since the physical volume
 is somewhat small, one should check the size of
 finite size corrections. We therefore performed a
 low statistics run (about $120$ gauge field configurations)
 for this $\beta$. The final result is also tabulated in
 Table~\ref{tab:mass_parameters} labelled as $\beta=2.8(II)$.
 We see that the physical bare quark mass parameters changes,
 especially for the strange quark. However, the optimal value
 for the parameter $\nu$ at physical quark masses remain
 compatible within one or two standard deviations.
 Therefore, for the purpose of tuning the parameter $\nu$,
 this result seems to indicate that the finite size
 corrections are not large. Of course, if one would use
 the action to calculate physical quantities and compare
 with the experimental values, it would be safer to use
 larger volumes, which is what we will do in the future.

\section{Conclusions}

 In this paper, we present a systematic numerical analysis
 on the tuning of the bare speed of light parameter $\nu$
 in the tadpole improved anisotropic Wilson quark action.
 The tuning is done in a quark mass dependent way with
 quark mass values ranging from the strange to the charm.
 The optimal values of $\nu$ are obtained for various
 values of $\beta$ using the pseudo-scalar meson
 dispersion relations. With the help of the anisotropic lattices
 with asymmetric volumes, the dispersion relations can be
 measured with good accuracy.
 Using the tadpole improved anisotropic Wilson action with
 these optimized parameters, a quenched calculation can then
 be performed to study properties of hadrons
 made up of either light or heavy quarks. Therefore, with
 the same improved quark action one can study hadron
 spectrum and other physical properties in a wide
 range of quark masses. We hope to come to this
 issue in the near future.

 \section*{Acknowledgments}

 We would like to thank Prof. H.~Q.~Zheng and
 Prof. S.~L.~Zhu of Peking University for helpful
 discussions. Our thanks also goes to Dr. J.P.~Ma
 at Institute of Theoretical Physics, Academia Sinica
 and Dr. Y.~Chen at Institute of High Energy Physics,
 Academia Sinica for their stimulating discussions.


\end{document}